\documentclass[epj]{svjour}
\usepackage{latexsym}
\usepackage{graphics}
\usepackage{amsfonts}
\usepackage{amssymb}
\usepackage{amsmath}
\begin{document}
\title{Cosmological Evolution of Interacting Dark Energy in Lorentz Violation}
\author{Freddy P. Zen\inst{1,2}, Arianto\inst{1,2,3}, Bobby E. Gunara\inst{1,2}, Triyanta\inst{1,2} \and A. Purwanto\inst{1,4}
}                     
%
%
\institute{Theoretical Physics Lab., THEPI Devision \and Indonesia Center for Theoretical and Mathematical Physics (ICTMP)\\
Faculty of Mathematics and Natural Sciences,
 Institut Teknologi Bandung,\\
Jl. Ganesha 10 Bandung 40132, INDONESIA. \and Department of Physics, Udayana University\\
Jl. Kampus Bukit Jimbaran Kuta-Bali 80361, INDONESIA \and Department of Physics, Institut Teknologi Sepuluh November\\
Jl. Kampus Sukolilo Surabaya 60111, INDONESIA. \\
\email{fpzen@fi.itb.ac.id,arianto@upi.edu,bobby@fi.itb.ac.id,\\
triyanta@fi.itb.ac.id,purwanto@physics.its.ac.id}}

\date{Received: date / Revised version: date}
%
\abstract{ The cosmological evolution of an interacting scalar
field model in which the scalar field interacts with dark matter,
radiation, and baryon via Lorentz violation is investigated. We
propose a model of interaction through the effective coupling
$\bar{\beta}$. Using dynamical system analysis, we study the
linear dynamics of an interacting model and show that the dynamics
of critical points are completely controlled by two parameters.
Some results can be mentioned as follows. Firstly, the sequence of
radiation, the dark matter, and  the scalar field dark energy
exist and baryons are sub dominant. Secondly, the model also
allows the possibility of having a universe in the phantom phase
with constant potential. Thirdly, the effective gravitational
constant varies with respect to time through $\bar{\beta}$. In
particular, we consider a simple case where $\bar{\beta}$ has a
quadratic form and has a good agreement with the modified
$\Lambda$CDM and quintessence models. Finally, we also calculate
the first post--Newtonian parameters for our model.
} 
\titlerunning{Interacting Dark Energy in Lorentz Violation}
\authorrunning{Freddy P. Zen et al}
\maketitle
\section{Introduction}
\label{intro} There has been a growing appreciation of the
importance of the violations of Lorentz invariance in recent
years. The intriguing possibility of the Lorentz violation is that
an unknown physics at high-energy scales could lead to a
spontaneous breaking of Lorentz invariance by giving an
expectation value to certain non Standard Model fields that carry
Lorentz indices, such as vectors, tensors, and gradients of scalar
fields~\cite{Kostelecky:1988zi}. Recently, it has been proposed a
relativistic theory of gravity where gravity is mediated by a
tensor, a vector, and a scalar field, thus called TeVeS
gravitational theory~\cite{Bekenstein:2004}. It provides modified
Newtonian dynamics (MOND) and Newtonian limits in the weak field
nonrelativistic limit, and is devoid of a causal propagation of
perturbations. TeVeS could also explain the large-scale structure
formation of the Universe without recurring to cold dark matter
\cite{Skordis:2006}, which is composed of very massive slowly
moving and weakly interacting particles. On the other hand, the
Einstein--Aether theory \cite{Jacobson:2000xp} is a vector-tensor
theory, and TeVeS can be written as a vector-tensor theory which
is the extension of the Einstein--Aether
theory~\cite{Zlosnik:2006}. In the case of generalized
Einstein--Aether theory~\cite{Zlosnik:2007}, the effect of a
general class of such theories on the solar system has been
considered in Ref.~\cite{Bonvin:2008}. Moreover, as has been shown
by authors in Ref.~\cite{Baojiu:2007}, the Einstein--Aether theory
may lead to significant modifications of the power spectrum of
tensor perturbation. The strong gravitational cases including
black holes of such theories have been studied in Refs.~\cite{BH}.

The existence of vector fields in a scalar-vector-tensor theory of
gravity also leads to its applications in modern cosmology and it
might explain inflationary scenarios \cite{Lim:2004js,KS} and
accelerated expansion of the universe
\cite{Zlosnik:2007,Tartaglia:2006;2007}. The accelerated expansion
and crossing of the phantom divided line has been studied recently
by authors in Ref.~\cite{Nozari:2008ff} Based on a dynamical
vector field coupled to the gravitation and scalar fields, we have
studied to some extent the cosmological implications of a
scalar-vector-tensor theory of gravity \cite{Arianto:2007}.

Since the discovery of accelerated expansion of our Universe
~\cite{Riessetal}, identifying the contents of dark energy and
dark matter is one of the most important subjects in modern
cosmology. The dark energy is described by an equation of state
parameter $\omega  = p/ \rho$, the ratio of the spatially
homogeneous dark energy pressure $p$ to its energy density $\rho$.
A value of $\omega < -1/3$ is required for accelerated expansion.
The classification of dark energy might be due to: quintessence
field~\cite{Ratraetal}, tachyon models~\cite{Senetal},  Chaplygin
gas~\cite{Yuetal} if $\omega>-1$, cosmological constant if
$\omega=-1$~\cite{Sahni:2000,Peebles:2003,Carroll:2001,Padmanabhan:2003},
or phantom field if $\omega<-1$~\cite{Caldwell:2002}. A recent
comprehensive review on dark energy is available
in~\cite{Copeland:2006}. Of course, as it has been discussed in
\cite{CAPicon:2004,Kiselev:2004} the vector field is also a viable
dark energy candidate and effects on the cosmic microwave
background radiation and the large scale structure
\cite{Koivisto:2005}.

In the previous work \cite{Arianto:2008}, the attractor solutions
in Lorentz violating scalar-vector-tensor theory of gravity
without interaction with background matter was studied. In this
framework, both the effective coupling and potential functions
determine the stabilities of the fixed points. In the model, we
considered the constants of slope of the effective coupling and
potential functions which lead to the quadratic effective coupling
with the (inverse) power-law potential. Differing from the
previous work, in this work, we investigate the cosmological
evolution of the scalar field dark energy and background perfect
fluid by means of dynamical system. We study the cases of scalar
field dark energy interacting with background perfect fluid. The
interaction terms are taken to be two different forms which are
mediated by the slope of the effective coupling. For more
realistic model we assume that the background matter fields might
be dark matter, radiation, and baryons.

Furthermore, to test the model in the solar system we present the
post--Newtonian parameters (PN). In the PN approximation we
restrict ourselves to the first post--Newtonian. The parameterized
post--Newtonian (PPN) parameters are determined by  expanding the
modified field equations in the metric perturbation. Then, we
compare the solution to the PPN formalism in first PN
approximation proposed by Will and
Nordtvedt~\cite{Will:1972zz,Will} and read off the coefficients
(the PPN parameters) of post Newtonian potentials of the theory.

This paper is organized as follows. In Section~\ref{sec:1}, we set
down the general formalism of the scalar field interacting with
background perfect fluid in the scalar-vector-tensor theory where
the Lorentz symmetry is spontaneously broken due to the unit-norm
vector field. We derive the governing equations of motion for the
canonical Lagrangian of the scalar field. In Section~\ref{sec:2},
we study the interaction models and the attractor solutions. The
critical points of the system and their stability are presented.
The cosmological implication is discussed in Section~\ref{sec:5}.
In Section~\ref{sec:6}, we present the parameterized
post-Newtonian parameters of the model. The final Section is
devoted to the conclusions.

In what follows, the conventions that we use throughout this work
are the following: Greek letters represent spacetime indices,
while Latin letters stand for spatial indices and repeated indices
mean Einstein's summation. The symbol ${\cal O}(N)$ stands for
terms of order $N$. Finally, we use the metric signature
$(-,+,+,+)$.
\section{The action and field equations}
\label{sec:1} In the present section, we develop the general
reconstruction scheme for the scalar-vector-tensor gravitational
theory. We will consider the properties of general
four-dimensional universe, i.e. the universe where the
four-dimensional space-time is allowed to contain any
non-gravitational degree of freedom in the framework of Lorentz
violating scalar-tensor-vector theory of gravity. Let us assume
that the Lorentz symmetry is spontaneously broken by imposing the
expectation values of a vector field $u^\mu$ as $<0| u^\mu u_\mu
|0> = -1$. The action can be written as the sum of four distinct
parts:
\begin{eqnarray}
     S&=& S_g + S_u + S_{\phi}+S_m \ ,
     \label{eq:action}
\end{eqnarray}
where the actions for the tensor field $S_g$, the vector field
$S_u$,  the scalar field $S_{\phi}$, and the ordinary matter
$S_{m}$, respectively, are given by
\begin{eqnarray}
    S_g &=& \int d^4 x \sqrt{-g}~ {1\over 16\pi G}R  \ ,
    \label{eq:act-tensor} \\
    S_u &=& \int d^4 x \sqrt{-g} \left[
  - \beta_1 \nabla^\mu u^\nu \nabla_\mu u_\nu-\beta_2 \left( \nabla_\mu u^\mu \right)^2
     \right. \nonumber\\
    &&-\beta_3 \nabla^\mu u^\nu \nabla_\nu u_\mu
    \left. + \lambda \left( u^\mu u_\mu +1 \right) \right]  \ ,
  \label{eq:act-vector} \\
  S_\phi &=&  \int d^4 x \sqrt{-g}~ \left[-{1\over 2}(\nabla \phi)^2 - V(\phi)\right] \ ,
   \label{eq:act-scalar} \\
   S_m &=& \int d^4 x \sqrt{-g}~ {\cal{L}}_m(\phi, \Psi_i, g_{\mu\nu}) \ .
   \label{eq:act-matter}
\end{eqnarray}
In the above $\beta_i(\phi)$ ($i=1,2,3$) are the functions of
$\phi$, and $\lambda$ is a Lagrange multiplier. In
Eq.~(\ref{eq:act-matter}), we allow for an arbitrary coupling
between the matter fields $\Psi$ and the scalar field $\phi$.

Varying the action (\ref{eq:action}) with respect to $g^{\mu\nu}$,
we have field equations
\begin{eqnarray}
   R_{\mu\nu}-{1\over 2}g_{\mu\nu}R = 8\pi G T_{\mu\nu} \ ,
   \label{eq:einstein-eq}
\end{eqnarray}
where $R_{\mu\nu}$ is the Ricci tensor, $R$ is the scalar
curvature, $g_{\mu\nu}$ is the metric tensor, and $T_{\mu\nu}$ is
the energy-momentum tensor for all the fields present, $T_{\mu\nu}
=T_{\mu\nu}^{(u)} + T_{\mu\nu}^{(\phi)} +T_{\mu\nu}^{(m)}$.
$T_{\mu\nu}^{(u)}$, $T_{\mu\nu}^{(\phi)}$ and $T_{\mu\nu}^{(m)}$
are the energy-momentum tensors of vector, scalar fields, and
ordinary matter, respectively, given by
\begin{eqnarray}
T^{(u)}_{\mu\nu}&=&2\beta_{1}\left(\nabla_{\mu}u^{\tau}\nabla_{\nu}u_{\tau}-\nabla^\tau
u_{\mu}\nabla_{\tau}u_{\nu}\right) -2\nabla_\tau\left(u_{(\mu}{J^\tau}_{\nu)}\right)\nonumber\\
&& -2\nabla_{\tau}\left(u^\tau
J_{(\mu\nu)}\right)+2\nabla_{\tau}\left(u_{(\mu}
{J_{\nu)}}^{\tau}\right)\nonumber\\
    &&- 2u_\sigma\nabla_\tau J^{\tau\sigma}u_\mu
u_\nu +g_{\mu\nu}{\cal L}_{u} \ ,
\label{eq:emvector} \\
T_{\mu\nu}^{(\phi)}&=& \nabla_\mu \phi \nabla_\nu \phi - {1\over
2}g_{\mu\nu} \left[ (\nabla\phi)^2 + 2 V(\phi)\right] \ ,
\label{eq:emscalar} \\
T_{\mu\nu}^{(m)} &=& (\rho_m + p_m)v_\mu v_\nu + p_m g_{\mu\nu} \
, \label{eq:emmatter}
    \end{eqnarray}
where $v^\mu$ is the four velocity and the current tensor
$J_{\mu\nu}$ in Eq.~(\ref{eq:emvector}) is given by
\begin{equation}
    J^\mu{}_\nu =-\beta_{1} \nabla^{\mu}u_{\nu}- \beta_{2} \delta^\mu_\nu \nabla_{\tau}u^{\tau}
    - \beta_{3}\nabla_{\nu}u^{\mu} \ .
    \label{eq:curtensor}
\end{equation}
The energy-momentum tensor $T_{\mu\nu}$ is conserved
\begin{equation}
    \nabla^\nu \left(T_{\nu\mu}^{(u)} + T_{\nu\mu}^{(\phi)}+T_{\nu\mu}^{(m)} \right)= 0 \ .
    \label{eq:bianchiiden}
\end{equation}
In general, however, the Bianchi identity implies that each energy
species in the cosmic mixture is not conserved, namely
\begin{eqnarray}
    \nabla^\nu T_{\nu\mu}^{(u)}= \sigma^{(u)}_\mu  \ , ~~
    \nabla^\nu T_{\nu\mu}^{(\phi)}= \sigma^{(\phi)}_\mu  \ , ~~
    \nabla^\nu T_{\nu\mu}^{(m)} = \sigma^{(m)}_\mu \ .
    \label{eq:interacmatter}
\end{eqnarray}
Here $\sigma^{(k)}_\mu$ ($k=u,\phi, m$) is an arbitrary vector
function of the space-time coordinates that determines the rate of
transfer of energy, where  $\sigma^{(u)}_\mu +\sigma^{(\phi)}_\mu
+\sigma^{(m)}_\mu =0$. This is in accordance with
Eq.~(\ref{eq:bianchiiden}). Equation~(\ref{eq:interacmatter}) is
the basic feature of interacting models in which there is exchange
of energy between the components of the cosmic fluid. Moreover,
the projection of the non conservation equation along the velocity
of the whole fluid $n^\mu$ is
\begin{eqnarray}
Q^{(u)} =-Q^{(\phi)} - Q^{(m)} \ , \label{eq:jmlinteract}
\end{eqnarray}
where $Q^{(k)}\equiv n^\mu \sigma^{(k)}_\mu$ is a scalar.

Using Eq.~(\ref{eq:einstein-eq}), the Friedmann and Raychaudhuri
equations can be written as
\begin{equation}
    3H^2= 8\pi G \left({\rho}_u +{\rho}_\phi +{\rho}_m  \right)  \ ,
    \label{eq:Friedmann}
\end{equation}
and
\begin{equation}
    2\dot{H}= -8\pi G \left({\rho}_u +{\rho}_\phi +{\rho}_m+{p}_u +{p}_\phi +{p}_m
    \right)\ ,
    \label{eq:Raychaudhuri}
\end{equation}
where
\begin{eqnarray}
    && \rho_u =  -3\beta H^2,\quad p_u =-\rho_u +2\left(\beta\dot{H}+ \dot{\beta}H\right),
    \label{eq:vectorcomp} \\
    && \rho_{\phi} = {1\over 2} \dot{\phi}^2 + V \ ,\quad p_{\phi}= -\rho_{\phi} + \dot{\phi}^2  \ .
    \label{eq:scalarcomp}
\end{eqnarray}
Here, we have defined $\beta \equiv \beta_1 +3 \beta_2 + \beta_3$.

Substituting Eqs.~(\ref{eq:vectorcomp}) and (\ref{eq:scalarcomp})
into Eqs.(\ref{eq:Friedmann}) and (\ref{eq:Raychaudhuri}),
respectively, we obtain
\begin{eqnarray}
    3\left( \beta + \frac{1}{8\pi G } \right) H^2=  {1\over 2} \dot{\phi}^2 + V + \rho_m \
    \label{eq:Friedmann1}
\end{eqnarray}
and
\begin{eqnarray}
    2 \left( \beta + \frac{1}{8\pi G } \right)\dot{H}=-2\dot{\beta}H
     - \dot{\phi}^2 - 2(\rho_m + p_m) \ .
    \label{eq:Raychaudhuri1}
\end{eqnarray}
Let us define the effective coupling as follows
\begin{eqnarray}
   \bar{\beta} &\equiv&   \beta + {1\over 8\pi G} \ ,
   \label{eq:twopart}
\end{eqnarray}
then Eqs.~(\ref{eq:Friedmann1}) and (\ref{eq:Raychaudhuri1}) can
be simplified as
\begin{eqnarray}
&&  H^2  = \frac{1}{3\bar{\beta}} \left( {1\over 2} \dot{\phi}^2 +
V + \rho_m \right) \ ,
\label{eq:Friedmann2}\\
&&  {\dot{H}\over H}=-{\dot{\bar{\beta}}\over \bar{\beta} }
     - {1\over 2}{\dot{\phi}^2 \over H\bar{\beta} } - \gamma_m{\rho_m \over H\bar{\beta} } \ .
\label{eq:Raychaudhuri2}
\end{eqnarray}
Here, we have defined $p_m=(\gamma_m-1)\rho_m$, where $\gamma_m$
is the ordinary matter barotropic parameter, which is related to
the equation of state parameter $\omega_m$ through the
relationship $\gamma_m=1+\omega_m$. Similarly, we also defined the
scalar field barotropic parameter,
$p_\phi=(\gamma_\phi-1)\rho_\phi$ and $\gamma_\phi=1+\omega_\phi$.
Then the effective equation of state for the total cosmic fluid is
\begin{eqnarray}
   \gamma^{(e)}=1+\frac{p_u+p_\phi+p_m}{\rho_u+\rho_\phi+\rho_m} \ ,
   \label{eq:eos-effective}
\end{eqnarray}
which, again, is related to the equation of state parameter
$\gamma^{(e)}$ through $\gamma^{(e)}=1+\omega^{(e)}$. The
condition for an accelerated universe is $\gamma^{(e)}<2/3$. When
$0<\gamma^{(e)}<2/3$, the universe is in quintessence phase while
it is in phantom phase when $\gamma^{(e)}<0$.

From Eq.~(\ref{eq:vectorcomp}) we obtain
\begin{eqnarray}
   \dot{\rho}_u + 3H({\rho}_u + p_u)=3H^2 \dot{\bar{\beta}}  \ .
   \label{eq:eos-vec}
\end{eqnarray}
In order to preserve the conservation of total energy equation
$\dot{\rho}_{tot} + 3H({\rho}_{tot} + p_{tot})=0$, where
$\rho_{tot} = \rho_u + \rho_{\phi}+ \rho_m$ and $p_{tot} =p_u +
p_{\phi}+ p_m$ are the total energy density and the pressure,
respectively, one can write the conservation of scalar field and
matter field:
\begin{eqnarray}
   &&\dot{\rho}_\phi + 3H({\rho}_\phi + p_\phi)=- 3H^2 \dot{\bar{\beta}}+Q_m \ ,
   \label{eq:concer-scalar}\\
   &&\dot{\rho}_m + 3H({\rho}_m + p_m)=Q_m \ .
   \label{eq:concer-matter}
\end{eqnarray}
The interaction term can be interpreted as a transfer from one
energy component to another energy component of the cosmic fluid.
These interactions are completely associated with Lorentz
violation. In our case, the scalar field decays into the matter
field and the vector field. The conservation of scalar field,
Eq.~(\ref{eq:concer-scalar}), is equivalent to a dynamical
equation for the scalar field $\phi$,
\begin{eqnarray}
Q_m=-\dot{\phi}\left(\ddot{\phi}+3H\dot{\phi} + V_{,\phi}+3H^2
\bar{\beta}_{,\phi}\right) \ . \label{eq:eosscalarde2}
\end{eqnarray}
The above equation reduces to Refs.~\cite{KS,Arianto:2008} for
$Q_m=0$. Equations~(\ref{eq:Friedmann2}),
(\ref{eq:Raychaudhuri2}), and (\ref{eq:eosscalarde2}) represent
the basic set of equations of the model of interacting components
of the cosmic fluid in the framework of Lorentz violating
scalar-vector-tensor theory of gravity. In what follows we shall
apply a dynamical system to analyze the cosmological dynamics of
this set of equations.
\section{Interacting model}
\label{sec:2} Some models that allow interaction between the
scalar field and the matter field have been proposed as a solution
to the cosmic coincidence problem. These models are compatible
with observational data but so far there has been no evidence on
the existence of this interaction. A solution will be achieved if
the dynamical system presents scaling solutions which are
characterized by a constant dark matter to dark energy ratio. Even
more important are those scaling solutions that are also
attractors and have the accelerated solution. In this way, the
coincidence problem gets substantially alleviated because,
regardless of the initial conditions, the system evolves towards a
final state where the ratio of dark matter to dark energy remains
constant.

The explicit form of Eq.~(\ref{eq:jmlinteract}) can be expressed
in the form
\begin{eqnarray}
Q_\phi+Q_m=-
\frac{\dot{\bar{\beta}}}{\bar{\beta}}(\rho_{\phi}+\rho_m) \ .
\end{eqnarray}
We assume the interaction term as follows
\begin{eqnarray}
~Q_m=\frac{\dot{\bar{\beta}}}{\bar{\beta}}\rho_{\phi}=-
\frac{\bar{\beta}_{, \phi}}{\bar{\beta}}\rho_m \dot{\phi}\  .
\label{interactmodel}
\end{eqnarray}
The interaction term (\ref{interactmodel}) means that the scalar
field can exchange energy with the background matter, through the
interaction between them. In this case the exchange energy is
mediated by the slope of the effective coupling.

Equations (\ref{eq:concer-scalar}) and (\ref{eq:concer-matter}),
respectively, become
\begin{eqnarray}
   &&\dot{\rho}_\phi + 3H({\rho}_\phi + p_\phi)=- \frac{{\bar{\beta}_{,\phi}}}{\bar{\beta}}\rho_{\phi}\dot{\phi} \ ,
   \label{eq:concer-scalar1}\\
   &&\dot{\rho}_m + 3H({\rho}_m + p_m)=- \frac{{\bar{\beta}_{,\phi}}}{\bar{\beta}}\rho_{m}\dot{\phi} \
   .
   \label{eq:concer-matter1}
\end{eqnarray}

For a more realistic model we assume that the matter fields might
be a combination of dark matter, $\rho_{c}$, radiation,
$\rho_{r}$, and baryons, $\rho_{b}$: $\rho_m =
\rho_{c}+\rho_{r}+\rho_{b}$. We also assume that the barotropic
equation of state for the radiation field $p_r = \rho_r/3$ and
that the baryons are non-relativistic particles so that $p_b = 0$
holds. Hence, the equations for the energy densities of radiation
and baryons are
\begin{eqnarray}
   \dot{\rho}_r+4H\rho_r=0 \ , \qquad \dot{\rho}_b+3H\rho_b=0 \ ,
\end{eqnarray}
respectively, and we find the well-known relationships: $\rho_r
\propto a^{-4}$ and $\rho_b \propto a^{-3}$, $a$ is a scale
factor. For the scalar field and the dark matter we have
\begin{eqnarray}
\dot{\rho}_{\phi} + 3H\gamma_{\phi}^{(e)}{\rho}_{\phi}=0 \ ,
\qquad
   \dot{\rho}_{c} + 3H\gamma_{c}^{(e)}{\rho}_{c}=0 \ ,
   \label{eq:eos-darkmatter}
\end{eqnarray}
where $\gamma_{\phi}^{(e)}$ and $\gamma_{c}^{(e)}$ are the
effective barotropic equation of state for scalar field and  dark
matter, respectively,
\begin{eqnarray}
\gamma_{\phi}^{(e)}=\gamma_{\phi}+\frac{\dot{\bar{\beta}}}{3H\bar{\beta}}
\ , \quad
   \gamma_{c}^{(e)}=1+ \frac{\dot{\bar{\beta}}}{3H\bar{\beta}}\left(1+\frac{\rho_r+\rho_b}{\rho_c} \right) \ .
   \label{eq:eos-darkmatter}
\end{eqnarray}
Notice that for $\dot{\bar{\beta}}/\bar{\beta}<0$ we have
$\gamma_\phi^{(e)}<\gamma_\phi$, $\gamma_c^{(e)}<\gamma_c$ and
both ${\rho}_\phi$ and ${\rho}_c$ with Lorentz violation will
dilute slower then that without Lorentz violation or
$\bar{\beta}=$ const. Thus $\dot{\bar{\beta}}/\bar{\beta}$ will
determine both the effective equations of state
$\gamma_\phi^{(e)}$ and $\gamma_c^{(e)}$.
\subsection{Dynamical analysis}
\label{sec:3} In order to study the dynamics of the model, we
shall introduce the following dimensionless
variables~\cite{Arianto:2007,Arianto:2008}:
\begin{eqnarray}
    &&x^2\equiv{\dot{\phi}^2 \over 6\bar{\beta}H^2} \ , \qquad \quad y^2 \equiv {V\over 3H^2\bar{\beta}} \ ,
    \label{def-xy}\\
    && \lambda_1 \equiv -{\bar{\beta}_{,\phi} \over \sqrt{\bar{\beta}}}\ ,
    \qquad \lambda_2 \equiv - \sqrt{\beta}{V_{,\phi}\over V}\ ,
    \label{def-lambda}\\
    && \Gamma_1 \equiv \frac{\bar{\beta} \bar{\beta}_{,\phi\phi}}{\bar{\beta}_{,\phi}^2} \ , \qquad
    \Gamma_2 \equiv \frac{VV_{,\phi\phi}}{V_{,\phi}^2} +{1\over
2}{\bar{\beta}_{,\phi}/\bar{\beta}\over V_{,\phi}/V} \ ,
    \label{def-gamma}
\end{eqnarray}
and, accordingly, the governing equations of motion could be
reexpressed as the following system of equations:
\begin{eqnarray}
H'&=& -{3\over 2}H\left(1+x^2-y^2+{1\over 3}z^2- \sqrt{6}\lambda_1
x  \right) \ ,
\label{auto-hubble}\\
x'&=&-x\left(3+{H'\over H}\right)\nonumber\\
&&    +\sqrt{3\over 2}\left(\lambda_1 + \lambda_2  \right)y^2
+2\sqrt{3\over 2}\lambda_1 x^2 \ ,
\label{auto-x-mat}\\
y' &=& -y\left({H'\over H}- \sqrt{3\over 2} (\lambda_1-\lambda_2)
x \right) \ ,
\label{auto-y-mat}\\
z'&=&-z \left(2+{H'\over H}-\sqrt{3\over 2} \lambda_1 x\right) \ ,
\label{auto-z-mat}\\
u'&=&-u \left({3\over 2}+{H'\over H}- \sqrt{3\over 2} \lambda_1
x\right) \ , \label{auto-u-mat}
\end{eqnarray}
where
\begin{eqnarray}
    z=\sqrt{\frac{\rho_r}{3\bar{\beta}H^2}}\ , \qquad u=\sqrt{\frac{\rho_b}{3\bar{\beta}H^2}}\ .
\end{eqnarray}
A prime denotes a derivative with respect to the natural logarithm
of the scale factor, $d/d\ln a = H^{-1}d/dt$. Equation
(\ref{eq:Friedmann2}) gives the following constraint equation:
\begin{eqnarray}
    \Omega_{c} =\frac{\rho_{c}}{3\bar{\beta}H^2}=1-x^2-y^2-z^2-u^2 \ ,
   \label{eq:costrainfrw}
\end{eqnarray}
where $\Omega_\phi=\rho_\phi/3\bar{\beta}H^2=x^2+y^2$,
$\Omega_r=\rho_r/3\bar{\beta}H^2=z^2$, and
$\Omega_b=\rho_b/3\bar{\beta}H^2=u^2$. Notice that $\Omega_i$,
$(i=\phi, c, r, b)$ are the effective cosmological density
parameters which are associated with the Lorentz violation.
\begin{table*}
\caption{Properties of the critical points.}
\label{table1}       
\begin{tabular}{lllll}
\hline\noalign{\smallskip}
Point&$(x,y,z,u)$&$\Omega_{\phi}$&$\gamma_{\phi}$&$\gamma_{eff}$  \\
\noalign{\smallskip}\hline\noalign{\smallskip}
$A_{+}$&$(+1,0,0,0)$&$1$&$2$&$2-2\sqrt{2\over 3}\lambda_1$\\

$A_{-}$&$(-1,0,0,0)$&$1$&$2$&$2+2\sqrt{2\over 3}\lambda_1$\\

$B$&$\left(\frac{\lambda_1+\lambda_2}{\sqrt{6}},\sqrt{1-\frac{(\lambda_1+\lambda_2)^2}{6}},0,0\right)$
&$1$&${(\lambda_1+\lambda_2)^2\over 3}$&$-{(\lambda_1^2-\lambda_2^2)\over 3}$\\

$C_{r}$&$\left(\sqrt{2\over 3\lambda_1^2},0,\sqrt{3-{2\over \lambda_1^2}},0\right)$&${2\over 3\lambda_1^2}$&$2$&${2\over 3}$\\

$D$&$\left(\frac{\sqrt{3/2}}{(\lambda_1+\lambda_2)},\frac{\sqrt{3/2}}{(\lambda_1+\lambda_2)},0,0\right)$
&$\frac{3}{(\lambda_1+\lambda_2)^2}$&$1$&$1-\frac{2\lambda_1}{\lambda_1+\lambda_2}$\\

$D_r$&$\left(\frac{2\sqrt{2/3}}{\lambda_2},\sqrt{\frac{4(\lambda_2-2\lambda_1)}{3\lambda_2^2(\lambda_1+\lambda_2)}},\sqrt{\frac{(4\lambda_1+\lambda_2)(\lambda_1\lambda_2+\lambda_2^2-4)}{\lambda_2^2(\lambda_1+\lambda_2)}},0\right)$
&$\frac{4}{\lambda_2(\lambda_1+\lambda_2)}$&$\frac{4(\lambda_1+\lambda_2)}{3\lambda_2}$&${4\over 3}\left({\lambda_2-\lambda_1\over \lambda_2}\right)$\\

$E_r$&$\left(0,0,1,0\right)$&$0$&$-$&${4\over 3}$\\

$E_b$&$\left(0,0,0,u_c\right)$&$0$&$-$&$1$\\
\noalign{\smallskip}\hline
\end{tabular}
\end{table*}
\begin{table}
\caption{Stabilities and acceleration conditions of the critical
points.}
\label{table2}       
\begin{tabular}{llll}
\hline\noalign{\smallskip}
Point&Existence&Stability&Acceleration  \\
\noalign{\smallskip}\hline\noalign{\smallskip}
$A_{+}$&$\forall \lambda_1,\lambda_2$&unstable&$\lambda_1>\sqrt{2\over 3}$\\

$A_{-}$&$\forall \lambda_1,\lambda_2$&unstable&$\lambda_1<-\sqrt{2\over 3}$\\

$B$&$(\lambda_1+\lambda_2)^2<6$&stable&$\lambda_2^2<\lambda_1^2+2$\\

$C_{r}$&$\lambda_1^2>{2\over 3}$&unstable&never\\

$D$&$(\lambda_1+\lambda_2)^2>3$&stable&$\lambda_2<5\lambda_1$\\

$D_r$&$\lambda_2(\lambda_1+\lambda_2)>4$&unstable&$\lambda_2<2\lambda_1$\\

$E_r$&$\forall \lambda_1,\lambda_2$&unstable&never\\

$E_b$&$\forall \lambda_1,\lambda_2$&unstable&never\\
\noalign{\smallskip}\hline
\end{tabular}
\end{table}
In general, the parameters $\lambda_1$, $\lambda_2$, $\Gamma_1$
and $\Gamma_2$ are  variables dependent on $\phi$ and completely
associated with the Lorentz violation. In order to construct
viable Lorentz violation model, we require that the effective
coupling $\bar{\beta}$ and the potential function $V$ should
satisfy the conditions $\Gamma_1 >1/2$ and $\Gamma_2
>1-\lambda_1/2\lambda_2$, respectively. In this paper, we want to discuss the phase
space, then we need certain constraints on the effective coupling
and potential function. Note that for $\beta_i=$ const.,
$\lambda_1\rightarrow 0$, the scalar field dynamics in the Lorentz
violating scalar-vector-tensor theories is then reduced to the
scalar field dynamics in the conventional one. But, the effective
gravitational constant is rescaled by Eq.~({\ref{eq:twopart}}). In
this case, the cosmological attractor solutions can be studied
through a scalar exponential potential of the form $V(\phi) =
V_0\exp(-\lambda_2\phi/\sqrt{\bar{\beta}})$ where $\bar{\beta}=$
const. This exponential potential gives rise to scaling solutions
for the scalar field \cite{Copeland:1998}. In this paper we
consider the case in which $\lambda_1$ and $\lambda_2$ are
constant parameters. For example, a constant $\lambda_1$ is given
by an effective coupling $\bar{\beta}=\xi \phi^2$ and we have
$\lambda_1=-2\sqrt{\xi}$. A constant $\lambda_2$ can only be
obtained as a combination of $\bar{\beta}(\phi)$ and $V(\phi)$,
one finds
\begin{eqnarray}
    V(\phi)=V_0 (\bar{\beta}(\phi))^s \ ,
    \label{eq:potfunction}
\end{eqnarray}
where $s=\lambda_2/\lambda_1$ is a constant parameter. In general,
one can write the potential as a function of effective coupling,
$V(\phi)\equiv f(\bar{\beta}(\phi))$.
\begin{figure*}
\resizebox{0.75\textwidth}{!}{%
  \includegraphics{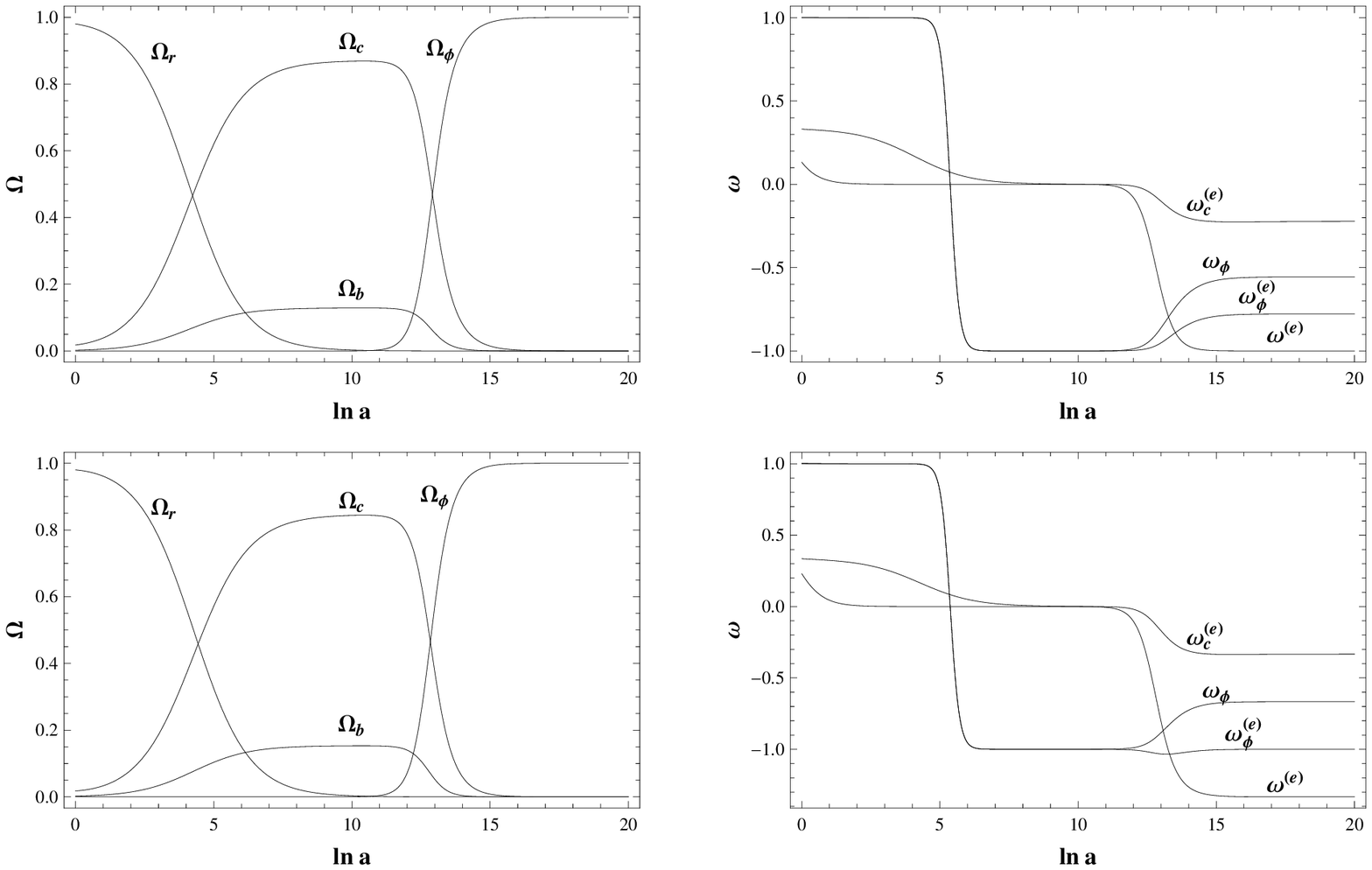}
} \caption{Evolution of the density parameters and the equation of
state parameters as a function of $\ln a$. Top panel corresponds
to the case of $\lambda_2=\lambda_1=-1/\sqrt{3}$ while the bottom
panel corresponds to the cases of constant potential and
$\lambda_1=-1$.}
\label{figure1}       
\end{figure*}
\begin{figure*}
\resizebox{0.75\textwidth}{!}{%
  \includegraphics{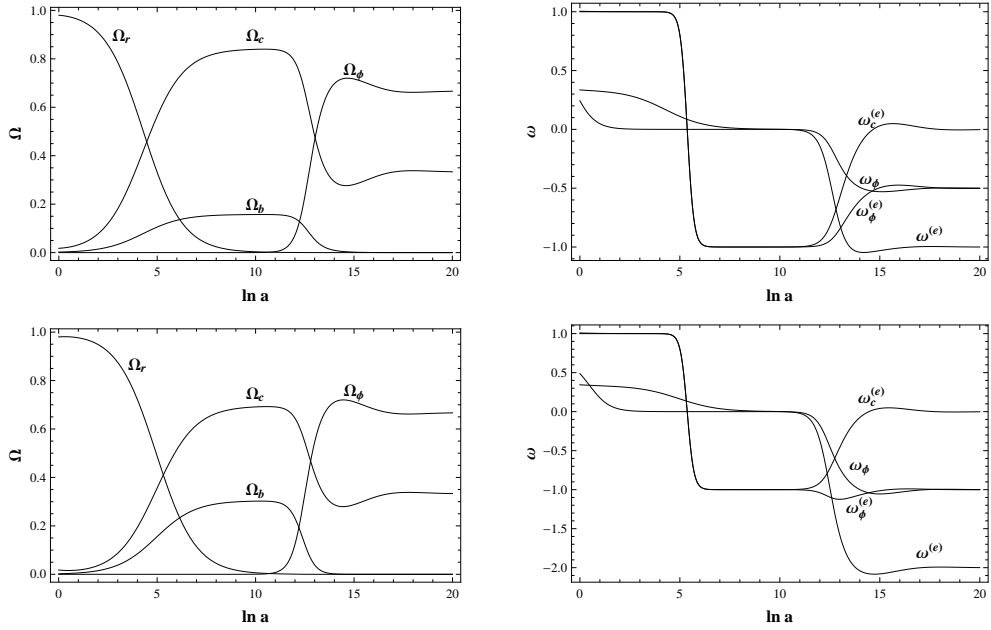}
} \caption{Evolution of the density parameters and the equation of
state parameters as a function of $\ln a$. Top panel corresponds
to the case of $\lambda_2=\lambda_1=-3/\sqrt{2}$ while the bottom
panel corresponds to the cases of constant potential and
$\lambda_1=3/2\sqrt{2}$.}
\label{figure2}       
\end{figure*}
\subsection{Attractor solutions}
\label{sec:4} The critical points $(x_c, y_c, z_c, u_c)$ are
obtained by imposing the conditions
$x^{\prime}=y^{\prime}=z^{\prime}=u^{\prime}=0$. Substituting
linear perturbation $x \rightarrow x_c+\delta x$, $y \rightarrow
y_c+\delta y$, $z \rightarrow z_c+\delta z$ and $u \rightarrow
u_c+\delta u$ about the critical points into
Eqs.~(\ref{auto-x-mat})--(\ref{auto-u-mat}), we obtain, up to
first-order in the perturbation, the equation of motion
\begin{eqnarray}
\label{matrix} \frac{d}{d\alpha}\left(
  \begin{array}{c}
    \delta x\\
    \delta y \\
    \delta z \\
    \delta u \\
  \end{array}
\right)= M\left(
  \begin{array}{c}
    \delta x \\
    \delta y \\
    \delta z \\
    \delta u \\
  \end{array}
\right) \ .
\end{eqnarray}
Notice from (\ref{auto-x-mat})--(\ref{auto-u-mat}) that the
dynamical equations are invariant under the change of sign
$(y,z,u) \rightarrow (-y,-z,-u)$, and in consequence we don´t have
to include the points with $(y,z,u)<0$ in our analyzes. The
properties of the critical points are summarized in
Table~\ref{table1}. There are eight critical points at all and two
of them lead to attractor solutions, depending on the values of
the parameters $\lambda_1$ and $\lambda_2$. The scalar field
dominated solution, point $B$ in Table~\ref{table2}, are
characterized by $\Omega=1$, and the effective equations of state
are given by
\begin{eqnarray}
\gamma_{\phi}^{(e)}={1\over 3}(\lambda_1+\lambda_2)^2\ , \quad
\gamma^{(e)}=-{1\over 3}(\lambda_1^2-\lambda_2^2) \ .
\label{eq:eff-scdominated1}
\end{eqnarray}

The solution of this point exists for $(\lambda_1+\lambda_2)^2<6$
and the universe is accelerated for $\lambda_2^2<\lambda_1^2+2$.
From eq.~(\ref{eq:eff-scdominated1}) one can see that the de
Sitter epoch corresponds to $\lambda_2=\lambda_1$.  The scalar
field is dark energy when $\lambda_1^2<1/2$. In this case the
effective coupling $\bar{\beta}$ and the potential function are
quadratic in $\phi$, $\bar{\beta}(\phi)\sim V (\phi)\sim \phi^2$.
The inflationary solution of this model has been studied in
Ref.~\cite{KS}. Figure \ref{figure1} shows that the sequence of
radiation, dark matter and  scalar field dark energy. The baryon
is sub--dominant in this case. The parameters correspond to
$\lambda_2=\lambda_1$ and $\lambda_1=-1/\sqrt{3}$. The scalar
field equation of state parameter $\omega_{\phi}=\gamma_{\phi}-1$
is nearly a constant, during the radiation and matter epochs
because the fields are almost frozen for which
$\omega_{\phi}=\omega_{\phi}^{(e)}$. At the transition era from
matter domination to the scalar field dark energy domination,
$\omega_{\phi}$ and $\omega_{\phi}^{(e)}$ begin to grow because
the kinetic energies of the fields become important. However, the
universe enters the de Sitter phase during which the field $\phi$
rolls up the potential. More interesting of this attractor
solution is of the constant potential, $\lambda_2=0$. The universe
is in phantom phase in this case because  $\omega^{(e)}$ is
crossing $-1$ and it is accelerated for $\lambda_1^2>-2$.

The second attractor solution is the scalar field scaling
solution, point $D$ in Table~\ref{table2}. The solution of this
point exists for $(\lambda_1+\lambda_2)^2>3$, corresponding to
energy density parameter
$\Omega_{\phi}=3/(\lambda_1+\lambda_2)^2$. The effective equations
of state are given by
\begin{eqnarray}
  &&\gamma_{\phi}=\gamma_m=1 \ , \quad \gamma_{\phi}^{(e)}=\gamma_{m}^{(e)}={\lambda_2 \over (\lambda_1+\lambda_2)}\ , \\
  && \gamma^{(e)}=1-\frac{2\lambda_1}{\lambda_1+\lambda_2} \ .
   \label{eq:eff-scalingdominated1}
\end{eqnarray}
The universe is accelerated for $\lambda_2<5\lambda_1$. In the
case of the effective coupling $\bar{\beta}$ and the potential
function are quadratic in $\phi$, i.e. $\lambda_2=\lambda_1$, the
universe is always accelerated. For the constant potential,
$\lambda_2=0$, the scalar field behaves as a cosmological constant
while the universe is in phantom phase. Figure \ref{figure2} shows
the sequence of radiation, dark matter and  scalar field dark
energy. The baryon is sub dominant in this case. The parameters
correspond to $\lambda_2=\lambda_1=-3/\sqrt{2}$ (top panel), and
$\lambda_1=3/2\sqrt{2}$ (bottom panel).
\section{A comparison of the model using supernova data}
\label{sec:5} From the above detail analysis, we may investigate
the cosmological consequences of a Lorentz violating
scalar-vector-tensor theory which incorporates time variations in
the gravitational constant. It was raised by Dirac who introduced
the large number hypothesis~\cite{Dirac}, and has recently become
a subject of intensive experimental and theoretical
studies~\cite{Uzan}. The effective gravitational constant,
$G^{(e)}$, is obtained from the Friedmann equation,
\begin{eqnarray}
G^{(e)}=\frac{1}{8\pi \bar{\beta}}=\frac{G}{1+8\pi G\beta} \ ,
\label{gcosmology}
\end{eqnarray}
where $G$ is the parameter in the action~(\ref{eq:action}).
Therefore the time variation of $G^{(e)}$ can be written as
\begin{eqnarray}
\frac{\dot{G}^{(e)}}{
G^{(e)}}=-\frac{\dot{\bar{\beta}}}{\bar{\beta}}\ , \label{varG}
\end{eqnarray}
and the effective gravitational constant is determined by the
effective coupling $\bar{\beta}$. For the quadratic effective
coupling, $\bar{\beta} \propto \phi^2$, the effective
gravitational constant is inversely proportional to $\phi^2$,
$G^{(e)} \propto [\phi(t)]^{-2}$. Recently using the data provided
by the pulsating white dwarf star G117-B15A the
astereoseismological bound on $\dot{G}/G$ is
found~\cite{Benvenuto} to be $-2.5 \times
10^{-10}~{yr^{-1}}<\dot{G}/G<4.0 \times 10^{-10}~{yr^{-1}}$.
\begin{figure*}
\resizebox{0.75\textwidth}{!}{%
  \includegraphics{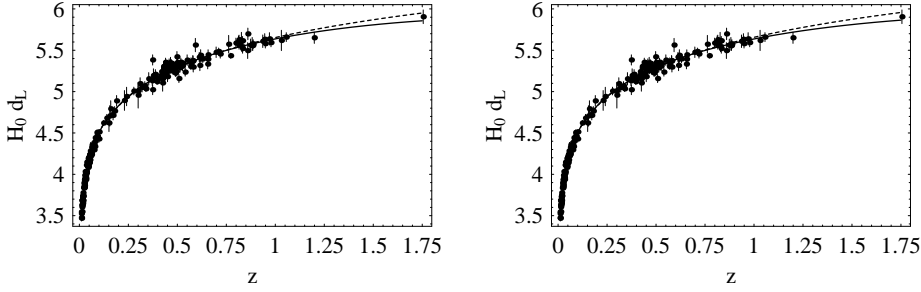}
} \caption{Observational 194 SnIa Hubble free luminosity distances
fitted  to our model. Left panel corresponds to the case of the
cosmological constant. The best fit values are $\zeta=-0.33$,
$\Omega_{m0}=0.24$. Right panel corresponds to the case of the
quintessence with constant equation of state parameter. The best
fit values are $\zeta=-0.29$, $\omega_{\phi}=-1.13$. Continuous
line denotes the curve in the context of Lorentz violating
scalar-vector-tensor theory, while dashed line denotes the
standard one.}
\label{figure3}       
\end{figure*}
In the present model the time variation in the gravitational
constant is given by
\begin{eqnarray}
\frac{\dot{G}^{(e)}}{ G^{(e)}}=\frac{3\lambda_1
}{(\lambda_1+\lambda_2)}H \ , \label{varg}
\end{eqnarray}
in the scaling solution and
\begin{eqnarray}
\frac{\dot{G}^{(e)}}{ G^{(e)}}=\lambda_1(\lambda_1+\lambda_2)H\ ,
\label{varg}
\end{eqnarray}
in the scalar field dominated solution, where the evolution of the
Hubble parameter is given by Eq.~(\ref{auto-hubble}). For
instance, in the case of power law expansion of the universe $a(t)
\propto t^p$ with $p>0$, the time variation of $G^{(e)}$ leads to
\begin{eqnarray}
\frac{\dot{G}^{(e)}}{ G^{(e)}}\propto\frac{3\lambda_1
}{(\lambda_1+\lambda_2)}t^{-1} \ , \label{varg1}
\end{eqnarray}
in the scaling solution. Assuming the present age of the Universe
as 14 Gyr, it is straightforward to derive from Eq.~(\ref{varg1})
the estimate $\dot{G}^{(e)}/G^{(e)}\sim 2.14 \times
10^{-10}~{yr^{-1}}$ for the case of constant potential. Our model
also allows the negative value of $\dot{G}^{(e)}/G^{(e)}$. Let us
focus on the scaling solution. If $\Omega_{\phi}=2/3$ we find
\begin{eqnarray}
\frac{\dot{G}^{(e)}}{ G^{(e)}}=\pm \sqrt{2} \lambda_1 H \ .
\label{varg2}
\end{eqnarray}
A negative $\dot{G}^{(e)}/G^{(e)}$ implies a time-decreasing
$G^{(e)}$, while a positive $\dot{G}^{(e)}/G^{(e)}$ means
$G^{(e)}$ is growing with time. From Eq.~(\ref{varg2}), it is
clear that the Lorentz violation leads to time variation of the
gravitational constant.

In the following, we study the expansion history of the universe
using the 194 SnIa data~\cite{Tonry:2003zg,Barris:2003dq}. We
simplify our model by considering an interaction between dark
matter and the scalar field dark energy given by
Eqs.~(\ref{eq:concer-scalar1}) and (\ref{eq:concer-matter1}). The
evolution of the dark matter and scalar field dark energy are
given by
\begin{eqnarray}
\rho_i(z)=\rho_{i0} e^{3\int_0^z
\frac{1+\omega_i^{(e)}(z')}{1+z'}dz'} \ , \qquad (i=m, \phi) \ ,
\end{eqnarray}
where $z=1/a-1$ is the redshift. Using the above relation, the
Hubble parameter as a function of the redshift  can be written as
\begin{eqnarray}
H^2(z) &=&
\left(\frac{H_0\bar{\beta}_0}{\bar{\beta}(z)}\right)^2\left[\Omega_{m0}(1+z)^3\right.
\nonumber\\
&&\left.+(1-\Omega_{m0})(1+z)^{3(1+\omega_{\phi}(z))} \right] \ ,
\end{eqnarray}
where the subscript $0$ describes the current value of the
variable. Notice that the evolution of the Hubble parameter is
deviated by the factor of $(\bar{\beta}_0/\bar{\beta})^{2}$
compared to the standard one. If the functions $\bar{\beta}(z)$
and $\omega_{\phi}(z)$ are given, we can find the evolution of the
Hubble parameter. In this section, we consider an ansatz for the
effective coupling,
\begin{eqnarray}
\bar{\beta}=\bar{\beta}_0 \left( 1 + \zeta z^2 \right) \ ,
\end{eqnarray}
where $\zeta$ is a constant.

Let us first consider the modified $\Lambda$ Cold Dark Matter
($\Lambda$CDM) model. We have
\begin{eqnarray}
H^2(z;\zeta,\Omega_{m0}) = \left(\frac{H_0}{1 + \zeta
z^2}\right)^2\left[\Omega_{m0}(1+z)^3+(1-\Omega_{m0}\right] \ .
\label{lcdm}
\end{eqnarray}
Equation (\ref{lcdm}) has two free parameters $\zeta$ and
$\Omega_{m0}$ which are determined by minimizing
\begin{eqnarray}
\chi^2=\sum_i \frac{\left[\mu_{obs}(z_i)-\mu(z_i)
\right]^2}{\sigma_i} \ ,
\end{eqnarray}
where $\mu$ is the extinction-corrected distance modulus,
\begin{eqnarray}
\mu(z) = 5 \log_{10} \left( \frac{d_L(z)}{1 Mpc}\right) +25 \ ,
\end{eqnarray}
and $\sigma_i$ is the total uncertainty in the SnIa data. The
luminosity distance is given by
\begin{eqnarray}
d_L(z) = {c(1+z) \over H_0} \int_0^z \frac{dz'}{H(z')}\ .
\end{eqnarray}

Fitting the model to 194 SnIa data, we get $\chi^2_{min}=195.68$,
$\zeta=-0.33$, and $\Omega_{m0}=0.24$. For comparison, we also fit
the cosmological constant model to the 194 SnIa data and find
$\chi^2=198.74$, and $\Omega_{m0}=0.34$.

In the next model we replace the cosmological constant energy
density by a scalar field dark energy with constant equation of
state parameter ($\omega_{\phi}(z)=$ constant). We set here
$\Omega_{m0}=0.3$. We evaluate $\chi^2(\zeta, \omega_{\phi})$ and
minimize with respect to $\zeta$ and $\omega_{\phi}$. We find
\begin{eqnarray}
\chi^2_{min} = \chi^2(\zeta=-0.29, \omega_{\phi}=-1.13) = 195.71 \
.
\end{eqnarray}

Figure ~\ref{figure3} shows a comparison of the observed 194 SnIa
Hubble free luminosity distances along the predicted curves in the
context of Lorentz violating scalar-vector-tensor theory. We see
that the effect of Lorentz violation appears at $z>0.75$. We
define the reduced form of Hubble parameter compared to the
standard case as
\begin{eqnarray}
H_{red}^2 = \frac{H^2_{LV}-H^2_{std}}{H^2_{std}} \ ,
\end{eqnarray}
where
\begin{eqnarray}
H^2_{std}(z) &=&H_0^2\left[\Omega_{m0}(1+z)^3 \right.
\nonumber\\
&&\left. +(1-\Omega_{m0})(1+z)^{3(1+\omega_{\phi}(z))} \right] \ .
\end{eqnarray}
Thus the reduced form of Hubble parameter, due to the effect of
Lorentz violation, is
\begin{eqnarray}
H_{red}^2(z) = \left(\frac{\bar{\beta}_0}{\bar{\beta}(z)}\right)^2
-1  \ .
\end{eqnarray}

\section{Parameterized Post-Newtonian} \label{sec:6} In order to
confront the predictions of a given gravity theory with experiment
in the solar system, it is necessary to compute its PPN
parameters. The post-Newtonian approximation is based on the
assumptions of weak gravitational fields and slow motions. It
provides a way to estimate general relativistic effects in the
fully nonlinear evolution stage of the large scale cosmic
structures. The procedure of parameterizing our model is following
to that of Ref~\cite{Foster:2005dk}, in which the authors has
derived the PPN parameters in the frame of Einstein aether theory.

In the weak field approximation, we choose a system of coordinates
in which the metric can be perturbatively expanded around
Minkowski spacetime. The decomposition is as follows
\begin{equation}
g_{\mu\nu} = \eta_{\mu\nu} + h_{\mu\nu} \ ,
\end{equation}
where $\eta_{\mu\nu}$ is the Minkowski metric and $h_{\mu\nu}$ is
the metric perturbations and we take $|h_{\mu\nu}|<<1$.

The equations governing the perturbation $h_{\mu\nu}$ in the model
Eq.~(\ref{eq:action}) are found by computing the Einstein field
equations in the perturbative limit. The full field equations are
given by
\begin{eqnarray}
R_{\mu\nu} = 8\pi G
\left(T^{(m)}_{\alpha\beta}+T^{(\phi)}_{\alpha\beta}+T^{(u)}_{\alpha\beta}
\right)\left(\delta^{\alpha}_{\mu} \delta^{\beta}_{\nu} - {1\over
2}g_{\mu\nu}g^{\alpha\beta} \right). \label{eq:einstein-equiv}
\end{eqnarray}

We allow for an arbitrary coupling between the matter fields and
the scalar field $\phi$. We assume that the scalar field $\phi$ is
coupled to a barotropic perfect fluid with a coupling function
given by
\begin{eqnarray}
q(\phi)=  -{1\over \rho_m\sqrt{-g}} {\delta S_m \over \delta\phi}
\ , \label{eq:einstein-equiv}
\end{eqnarray}
Therefore, the equation of motion for the scalar field is
\begin{eqnarray}
\square \phi -{d V \over d\phi}-\sum_{i=1}^3 {d\beta_i\over
d\phi}K_i=q(\phi)\rho_m \ , \label{eq:eosscalar-field}
\end{eqnarray}
where
\begin{eqnarray}
K_1 = \nabla^\mu u^\nu \nabla_\mu u_\nu,~ K_2=\left(\nabla_\mu
u^\mu\right)^2,~ K_3=\nabla^\mu u^\nu \nabla_\nu u_\mu.
\end{eqnarray}
In the previous discussion we have considered the coupling between
the scalar field and the matter field is given by the effective
coupling, $q(\phi)=d\ln \bar{\beta}/d\phi$, where $\bar{\beta}$ is
defined by Eq.~(\ref{eq:twopart}).

We also have the vector field equation,
\begin{equation}
\nabla_\mu {J^{\mu}}_{\nu}=\lambda u_{\nu} \ , \label{eq:vector}
\end{equation}
where ${J^{\mu}}_{\nu}$ is given by Eq.~(\ref{eq:curtensor}). The
constraint for the vector field is
\begin{eqnarray}
g_{\mu\nu}u^{\mu} u^{\nu} +1 =0 \ . \label{eq:constrain}
\end{eqnarray}
The vector field is purely timelike at the zeroth order and the
fluid variables are assigned orders of $\rho\sim \Pi \sim p/\rho
\sim (v^i)^2 \sim {\mathcal O}(1)$. The scalar field is expanded
as $\phi=\phi_0+{\mathcal O}(1)$, where $\phi_0$ is determined by
the cosmological solution. Then, the metric perturbations
$h_{\mu\nu}$ will be of orders $h_{00}\sim {\mathcal
O}(1)+{\mathcal O}(2)$, $h_{ij}\sim {\mathcal O}(1)$, and
$h_{0i}\sim {\mathcal O}(1.5)$.

The general form of the first post Newtonian metric is given
by~\cite{Will}
\begin{eqnarray}
g_{00} &=& -1+2U-2\beta_{PPN} U^2 - 2\xi \Phi_W  -(\zeta_1-2\xi){\mathcal A}\nonumber\\
&&+(2\gamma _{PPN}+2 +\alpha_3+\zeta_1 -2\xi)\Phi_1 \nonumber\\
       &&+2(3\gamma_{PPN}-2\beta_{PPN} +1 +\zeta_2 + \xi)\Phi_2  \nonumber\\
       &&+ 2(1+\zeta_3)\Phi_3+2(3\gamma_{PPN}+3\zeta_4-2\xi)\Phi_4  \ , \nonumber\\
g_{ij} &=& (1+2\gamma_{PPN} U)\delta_{ij} \ , \nonumber\\
g_{0i} &=&  -{1\over 2} (4\gamma_{PPN} + 3 +\alpha_1-\alpha_2
+\zeta_1 - 2\xi)V_i \nonumber\\
&&- {1\over 2} (1 +\alpha_2 -\zeta_1 - 2\xi)W_i \ . \label{eq:PPN
metric}
\end{eqnarray}
The PPN potentials ($U, \Phi_W, \Phi_1, \Phi_2, \Phi_3, \Phi_4,
{\mathcal A}, V_i, W_i$) are defined by
\begin{equation}
F(x)=G_N \int {d^3 y \frac{\rho(y)f}{|x-y|}} \ ,
\end{equation}
where the correspondences $F : f$ are given by
\begin{eqnarray}
&&U : 1,\quad \Phi_1 : v_i v_i, \quad \Phi_2:U, \quad \Phi_3:\Pi, \quad \Phi_4:{P\over \rho_m} \nonumber\\
&&\Phi_W:\int{d^3 z \rho_m(z) \frac{(x-y)_j}{|x-y|^2}\left( \frac{(y-z)_j}{|x-z|}\right)}, \quad V_i:v^i, \nonumber\\
&&{\mathcal A}:\frac{(v_i(x-y)_i)^2}{|x-y|^2},\quad
W_i:\frac{v_j(x-y)_j (x-y)^i}{|x-y|^2} \ .
\label{eq:corrppotential}
\end{eqnarray}
These potentials satisfy the following relations
\begin{equation}
F_{,ii}=-4\pi G_N \rho_m f \ ,
\end{equation}
for $U$, $\Phi_{1,2,3,4}$, and $V_i$. The superpotential $\chi$ is
defined by
\begin{equation}
\chi = -G_N \int{d^3y \rho_m |x-y|} \ ,
\end{equation}
which satisfies
\begin{equation}
\chi_{,ii}=-2U \ .
\end{equation}
We also note the identity
\begin{equation}
\chi_{,0i}=V_i - W_i \ .
\end{equation}
The PPN metric Eq.~(\ref{eq:PPN metric}) contains ten parameters
($\gamma_{PPN}$, $\beta_{PPN}$, $\xi$, $\alpha_{1,2,3}$,
$\zeta_{1,2,3,4}$). The parameter $\gamma_{PPN}$ measures how much
space-curvature is produced by a unit rest mass, while the
parameter $\beta_{PPN}$ determines how much non-linearity is there
in the superposition law of gravity. On the other hand, the
parameter $\xi$ determines whether there are preferred-location
effects, while $\alpha_{1,2,3}$ represent preferred-frame effects.
Finally, the parameters  $\zeta_{1,2,3,4}$ measure the amount of
violation of conservation of total momentum. In terms of
conservation laws, one can interpret these parameters as a measure
whether a theory is fully conservative, i.e. the linear and
angular momenta are conserved ($\zeta_{1,2,3,4}$ and
$\alpha_{1,2,3}$ vanish), semi-conservative, i.e. the linear
momentum is conserved ($\zeta_{1,2,3,4}$ and $\alpha_{1,2,3}$
vanish), or nonconservative, where only the energy is conserved
through lowest Newtonian order. One can verify that in general
relativity $\gamma_{PPN}=\beta_{PPN}=1$  and all other parameters
vanish, which implies that there are no preferred-location or
frame effects and that the theory is fully conservative.

The PPN parameters are determined as follows: expand the modified
field equations in the metric perturbation and in the PN
approximation; iteratively solve for the metric perturbation to
${\mathcal O}(2)$ in $h_{00}$, to ${\mathcal O}(1.5)$ in $h_{0i}$
and to ${\mathcal O}(1)$ in $h_{ij}$; compare the solution to the
PPN metric of Eq.~(\ref{eq:PPN metric}) and read off the PPN
parameters of the theory.

The gauge condition we use for the metric is as
follows~\cite{Foster:2005dk}
\begin{eqnarray}
&& h_{ij,j}={1\over 2}(h_{jj,i}-h_{00,i}) \ ,
\label{gauge1}\\
&&h_{0i,i}= 3U_{,0}+ B u^{i}_{,i} \ ,\label{gauge2}
\end{eqnarray}
where $B$ is a function of $\beta_i$ which will be determined
below.

There are two important notes to be considered here since we
concern in the post-Newtonian expansion. Firstly, the term
$V(\phi_0)$ is the same order as the energy density of the
cosmological constant, therefore, these term cannot leads to any
observable deviations at Solar system scales. Practically, we can
assume $dV(\phi)/d\phi=0$ as far as the post-Newtonian expansion
is concerned. It means that the cosmological solution corresponds
to a minimum of the potential and the scalar field will satisfy
the extremum condition. Secondly, from the analysis of the tensor
perturbations, the velocity of the gravitational waves are
different from the velocity of light~\cite{KS}. In order to have
the real velocity, the coupling functions $\beta_1$ and $\beta_3$
must be satisfy $(\beta_1+\beta_3)<(16\pi G)^{-1}$. Therefore, one
can assume that $\beta_1$ and $\beta_3$ are constant. This
assumption does not affect our previous calculations because the
coupling functions $\beta_1$ and $\beta_3$ have included in the
definition of the effective coupling, which is given by
Eq.~(\ref{eq:twopart}). These two assumptions will be used in the
following calculations.

\subsection{Solving $u^{\mu}$, $\phi$, and $g_{\mu\nu}$ to ${\cal
O}(1)$}

We first solve the constraint equation (\ref{eq:constrain}) to
${\cal O}(1)$. We find
\begin{eqnarray}
    u^{0}=1+{1\over 2}h_{00} \ .
\end{eqnarray}
From this result, we have
\begin{eqnarray}
u_{0}= -1+{1\over 2}h_{00} \ ,\qquad u_{i}= u^i + h_{0i} \ ,
\end{eqnarray}
where we used $u_{\mu}=g_{\mu\nu}u^{\nu}$.

From Eq.~(\ref{eq:eosscalar-field}), we obtain
\begin{eqnarray}
\phi^{(1)}_{,ii} - q_0 \rho_m=0  \ .
\end{eqnarray}
Using $U_{,ii}=-4\pi G_N\rho_m$, we have
\begin{eqnarray}
    \phi^{(1)}_{,ii}=-{q_0 \over
    4\pi G_N }U_{,ii} \ .
\end{eqnarray}
The solution of this equation is
\begin{eqnarray}
    \phi^{(1)}=-{q_0 \over
    4\pi G_N }U =-{\ln^{\prime} \bar{\beta}_0 \over
    4\pi G_N }U \ ,
\end{eqnarray}
where $q_0=q(\phi_0)$ and $\ln^{\prime} \bar{\beta}_0=d\ln
\bar{\beta}(\phi_0)/d\phi$.

The general expression of the covariant derivatives of $u_\mu$ is
\begin{eqnarray}
\nabla_{\mu}u^{\nu}= u^{\nu}_{,\mu} + \frac{u^{0}}{2}
g^{\nu\alpha}\left(h_{0\alpha,\mu} + h_{\alpha\mu,0}
-h_{0\mu,\alpha} \right) \ . \label{gencovar}
\end{eqnarray}
To ${\cal O}(2)$, we find
\begin{eqnarray}
\nabla_{\mu}u^{0}= 0, \quad \nabla_{0}u^{i}=u_{i,0} - {1\over
2}h_{00,i} + {1\over 4}h_{00}h_{00,i} \ , \label{covar}
\end{eqnarray}
and
\begin{eqnarray}
\nabla_{i}(\nabla_{0}u^{i})=u_{i,0i} - {1\over 2}h_{00,ii} +
{1\over 4}h_{00}h_{00,ii} - {3\over 4}h_{00,i}h_{00,i}\ .
\label{cocovar}
\end{eqnarray}
We also have
\begin{eqnarray}
\nabla_{i}u^{j}=u_{j,i} + {1\over 2}\left(h_{ji,0} - h_{0j,i}
-h_{0i,j} \right) \ ,
\end{eqnarray}
to ${\cal O}(1.5)$.

Let us now proceed with the solution to the field equation for the
"time-time" component of the metric perturbation. The left hand
side of Eq.~(\ref{eq:einstein-equiv}) is
\begin{eqnarray}
R_{00} &=& -{1\over 2}h_{00,ii} + {1\over 2}h_{ij}h_{00,ij} +
\left(h_{i0,i} -{1\over 2}h_{ii,0} \right)_{,0} \nonumber\\
&& -{1\over 4}h_{00,i}h_{00,i} + {1\over 2} h_{00,j}\left(h_{ij,i}
-{1\over 2}h_{ii,j} \right) \ ,
\end{eqnarray}
to ${\cal O}(2)$. Thus, at ${\cal O}(1)$ we have
\begin{eqnarray}
R_{00} = -{1\over 2}h_{00,ii}  \ .
\end{eqnarray}
To ${\cal O}(1)$, we have the components of energy-momentum tensor
\begin{eqnarray}
T^{(m)}_{00} = \rho_m, \quad T^{(m)}_{ij} = 0, \quad
T^{(\phi)}_{00}=0= T^{(\phi)}_{ij}  \ ,
\end{eqnarray}
and
\begin{eqnarray}
T^{(u)}_{00} &=&
-2\nabla_{0}J_{00}-2\nabla_{\mu}J_{0}{}^{\mu}=-2\nabla_{i}J_{0}{}^{i}\nonumber\\
&=&2(\beta_1\nabla_{0}u^i)=- \beta_{10}h_{00,ii} \ ,  \\
T^{(u)}_{ij} &=&0 \ ,
\end{eqnarray}
here $\beta_{i0}=\beta_{i}(\phi_0)$, $i=1,2,3$. Then, we obtain
\begin{eqnarray}
    (1-8\pi G\beta_{10})h_{00,ii}=-8\pi G \rho_m \ ,
\end{eqnarray}
which gives $h_{00}$ to ${\cal O}(1)$,
\begin{eqnarray}
    h_{00}=2U \ ,
    \label{solmetrico1}
\end{eqnarray}
with Newton's constant
\begin{eqnarray}
    G_N = {G\over (1-8\pi G\beta_{10})}\ .
    \label{defNewton}
\end{eqnarray}

To ${\cal O}(1)$ the ''space-space'' component of left hand side
Eq.~(\ref{eq:einstein-equiv}) is
\begin{eqnarray}
R_{ij} &=& -{1\over 2}h_{ij,kk} -{1\over 2}h_{kk,ij}+ {1\over
2}h_{00,ij}\nonumber\\
&&+ {1\over 2}(h_{ki,jk} - h_{kj,ik})  \ .
\end{eqnarray}
If we use the gauge (\ref{gauge1}), we obtain
\begin{eqnarray}
R_{ij} &=& -{1\over 2}h_{ij,kk}  \ .
\end{eqnarray}
Then, we have
\begin{eqnarray}
    (1-8\pi G\beta_{10})h_{ij,kk}=-8\pi G \rho_m \delta_{ij}\ .
\end{eqnarray}
This expression similar to the ''time-time'' component. It means
that the spatial metric perturbation to ${\cal O}(1)$ is then
simply given by the GR prediction without correction, namely
\begin{eqnarray}
    h_{ij}=h_{00} \delta_{ij} \ ,
\end{eqnarray}
where $h_{00}$ is given by Eq.~(\ref{solmetrico1}).

\subsection{Solving $u^{i}, g_{0i}$ to ${\cal O}(1.5)$}

From the previous results one can see that the Lagrange multiplier
is $\lambda \sim {\cal O}(1)$. Therefore, the vector field
equation yields $\nabla_{\mu}J^{\mu}{}_{i}=0$. This means
\begin{eqnarray}
    -J_{0i,0}+J_{ji,j} = 0 \ ,
\end{eqnarray}
to ${\cal O}(1.5)$, where
\begin{eqnarray}
J_{0i,0}&=& -(\beta_1\nabla_0u_i)_{,0}\ , \\
J_{ji,j}&=&-(\beta_1\nabla_ju_i)_{,j}-(\beta_2\nabla_ku_k)_{,i}-(\beta_3\nabla_ju_i)_{,j}
\ .
\end{eqnarray}
We notice that $\nabla_0 u_i = -{1\over 2}h_{00,i}$ to ${\cal
O}(1)$. To ${\cal O}(1.5)$, we have
\begin{eqnarray}
J_{0i,0}= {1\over 2}\beta_{10}h_{00,i0} \ , \label{j0i0}
\end{eqnarray}
and
\begin{eqnarray}
J_{ji,j}&=& -
\beta_{10}u^{i}_{,jj}-(\beta_{20}+\beta_{30})u^{j}_{,ij}- (\beta_{10}-\beta_{30})h_{0[i,j]j}\nonumber\\
&&- {1\over 2} (\beta_{10}+3\beta_{20}+\beta_{30})h_{00,0i} \ .
\end{eqnarray}
Then, the vector field equation can be written
\begin{eqnarray}
&&\beta_{10}u^{i}_{,jj}+(\beta_{20}+\beta_{30})u^{j}_{,ij}+(\beta_{10}-\beta_{30})h_{0[i,j]j}\nonumber\\
&&- {1\over 2} (2\beta_{10}+3\beta_{20}+\beta_{30})\chi_{,0ijj}=0
\ . \label{solvingui}
\end{eqnarray}
By taking the spatial divergence of this equation, we can solve
for $u^{i}_{,i}$,
\begin{eqnarray}
u^{i}_{,i}=C_{0}\chi_{,0ii} \ , \label{solvinguii}
\end{eqnarray}
where
\begin{eqnarray}
C_{0} =
\frac{(2\beta_{10}+3\beta_{20}+\beta_{30})}{2(\beta_{10}+\beta_{20}+\beta_{30})}
\ .
\end{eqnarray}
Using Eq.~(\ref{solvinguii}), the gauge (\ref{gauge2}) can be
rewritten
\begin{eqnarray}
h_{0i,i} =-{1\over 2}(3-2BC_{0})\chi_{,0ii}  \ . \label{gauge21}
\end{eqnarray}
Substituting Eqs.~(\ref{solvinguii}) and (\ref{gauge21}) into
Eq.~(\ref{solvingui}), and using the previous results, we can
solve Eq.~(\ref{solvingui}) for $u^i$,
\begin{eqnarray}
u^{i} &=&-{(\beta_{10}-\beta_{0})\over 2\beta_{10}}h_{0i}
\nonumber\\
&& +\left[C_{0} -{(\beta_{10}-\beta_{0})(3-2BC_{0})\over
4\beta_{10}}\right]\chi_{,0i}  \ . \label{solvingui1}
\end{eqnarray}

Let us now look for solutions to the field equations for the
metric perturbation $g_{0i}$ to ${\cal O}(1.5)$. To ${\cal
O}(1.5)$, we have the ''time-space'' components of left hand side
Eq.~(\ref{eq:einstein-equiv}),
\begin{eqnarray}
R_{0i}= -{1\over 2}h_{0i,jj}+ {1\over 4}(1+2BC_{0}) \ ,
\label{tso15}
\end{eqnarray}
and the ''time-space'' components of $T_{\mu\nu}$,
\begin{eqnarray}
T^{(m)}_{0i} = -\rho_m v_i, \quad  T^{(\phi)}_{0i}=0 \ ,
\end{eqnarray}
and
\begin{eqnarray}
T^{(u)}_{0i} = -J_{0i,0}-J_{ij,j} \ ,
\end{eqnarray}
where $J_{0i,0}$ is given by Eq.~(\ref{j0i0}) and
\begin{eqnarray}
J_{ij,j}&=& -(\beta_{10}+\beta_{20})u^{j}_{,ij} - \beta_{30}u^{i}_{,jj}- (\beta_{10}-\beta_{30})h_{0[j,i]j}\nonumber\\
&&- {1\over 2} (\beta_{10}+\beta_{30})h_{ij,0j}- {1\over 2}
\beta_{20}h_{jj,0i} \ .
\end{eqnarray}
Using Eqs.~(\ref{gauge21}) and (\ref{solvingui1}), and our
previous results, we obtain
\begin{eqnarray}
T^{(u)}_{0i} &=& \frac{(\beta_{10}^2 - \beta_{30}^2)}{2\beta_{10}} h_{0i,jj} \nonumber\\
& -& \left[\beta_{10} - \frac{(\beta_{10}^2 -
\beta_{30}^2)(3-2B_{0}C_{0})}{4\beta_{10}} \right]\chi_{,0ijj}\ .
\end{eqnarray}
Here, we have assumed that $B$ is a constant parameter, $B
\rightarrow B_{0}$, which will be determined below.

By solving the ''time-space'' components of the field equation, we
obtain
\begin{eqnarray}
&& \left[1 -  \frac{8\pi G(\beta_{10}^2 - \beta_{30}^2)}{\beta_{10}}\right]h_{0i,jj}= 16\pi G\rho v_i \nonumber\\
&&\qquad\qquad - \left(16\pi GE_{0}-B_{0}C_{0} - {1\over
2}\right)\chi_{,0ijj} \ , \label{solvetscomp}
\end{eqnarray}
where
\begin{eqnarray}
E_{0}=\beta_{10} - \frac{(\beta_{10}^2 -
\beta_{30}^2)(3-2B_{0}C_{0})}{4\beta_{10}}  \ ,
\end{eqnarray}
We can thus solve Eq.~(\ref{solvetscomp}) for $h_{0i}$,
\begin{eqnarray}
&& h_{0i}=-\left[1 -  \frac{8\pi G(\beta_{10}^2
-\beta_{30}^2)}{\beta_{10}}\right]^{-1} \times \nonumber\\
&&\quad\quad \left\{ \left[4(1-8\pi G\beta_{10})+16\pi
GE_{0}-B_{0}C_{0} -
{1\over 2} \right]V_i \right. \nonumber\\
&&\quad\quad \left.- \left(16\pi GE_{0}-B_{0}C_{0} - {1\over
2}\right)W_i \right\} \ ,
\end{eqnarray}
where we have used that the superpotential $\chi$ satisfies
$\chi_{,0i}=V_i-W_i$.

\subsection{Solving $g_{00}$ to ${\cal O}(2)$}

A full analysis of the PPN parameters requires that we solve for
the ''time-time'' component of Eq.~(\ref{eq:einstein-equiv}) to
${\cal O}(2)$. To ${\cal O}(2)$ we have
\begin{eqnarray}
R_{00}=-{1\over 2}\left( \tilde{h}_{00}+ 2U +2U^2 - 8\Phi_2 -
2B_{0}C_{0}\chi_{,00} \right)_{,ii}, \label{r00o2}
\end{eqnarray}
where we have defined $\tilde{h}_{00}=g_{00} + 1 - 2U$. To ${\cal
O}(2)$, we also have
\begin{eqnarray}
&&T^{(m)}_{00} = \rho_m(1+\Pi + v_iv_i-2U),~~T^{(m)}_{ii} = \rho_m
v_iv_i + 3p_m, \nonumber\\
&&T^{(\phi)}_{00}= {1\over 2} (\phi^{(1)}_i)^2,~~
T^{(\phi)}_{ii}=-{1\over
2}(\phi^{(1)}_i)^2,\nonumber\\
&&T^{(u)}_{00}
=\beta_1(\nabla_{0}u^i)^2+2u_{0}\nabla_{i}J_{0}{}^{i}, \nonumber\\
&&T^{(u)}_{ii} =\beta_1(\nabla_{0}u^i)^2-2\nabla_{0}J_{ii} \ .
\label{emto2}
\end{eqnarray}
Using the covariant derivative Eqs.~(\ref{covar}) and
(\ref{cocovar}), the components of energy-momentum tensor for the
vector field become
\begin{eqnarray}
T^{(u)}_{00} &=&-\beta_{10}\left[\tilde{h}_{00} + 2U + {5\over 2}
U^2 - 9\Phi_2 \right]_{,ii} \nonumber\\
&&- \beta_{10}[3-2C_0(1+B_0)]\chi_{,00ii}\ ,
\end{eqnarray}
and
\begin{eqnarray}
T^{(u)}_{ii} &=&\beta_{10}\left({1\over 2}U^2 -
\Phi_2\right)_{,ii}
\nonumber\\
&&-(\beta_{10}+3\beta_{20} +\beta_{30}) (3-2C_0)\chi_{,00ii} \ .
\end{eqnarray}
From Eq.~(\ref{emto2}) one can evaluate the ''time-time''
components of the right hand side of
Eq.~(\ref{eq:einstein-equiv}),
\begin{eqnarray}
&&T^{(m)}_{00} - {1\over 2} g_{00} g_{\mu\nu}T^{(m)\mu\nu}={1\over
2}(T^{(m)}_{00}+T^{(m)}_{ii})  \nonumber\\
&&=-\frac{1-8\pi G\beta_{10}}{8\pi G}\left( U+ 2\Phi_1 -
2\Phi_2+\Phi_3+3\Phi_4\right),  \label{mattf} \\
&&T^{(\phi)}_{00} - {1\over 2} g_{00}
g_{\mu\nu}T^{(\phi)\mu\nu}={1\over
2}(T^{(\phi)}_{00}+T^{(\phi)}_{ii})=0, \label{scaf}
\end{eqnarray}
and
\begin{eqnarray}
&&T^{(u)}_{00} - {1\over 2} g_{00} g_{\mu\nu}T^{(u)\mu\nu}={1\over
2}(T^{(u)}_{00}+T^{(u)}_{ii})  \nonumber\\
&&=-{1\over 2}\beta_{10}\left[\tilde{h}_{00} + 2U + 2 U^2 -8\Phi_2
\right]_{,ii}+{1\over
2}[2B_0C_0\nonumber\\
&&-(2\beta_{10}+3\beta_{2}+\beta_{30})(3-2C_0)]\chi_{,00ii}.
\label{vecf}
\end{eqnarray}
Using the ''time-time'' component of Eq.~(\ref{eq:einstein-equiv})
to ${\cal O}(2)$ and combining Eqs.~(\ref{r00o2}), (\ref{mattf}),
(\ref{scaf}) and (\ref{vecf}), we can obtain
\begin{eqnarray}
\tilde{h}_{00}&=&-2U^2+ 4\Phi_1 + 4\Phi_2
 + 2\Phi_3 + 6\Phi_4 \nonumber\\
&&+ Q_0 \chi_{,00} \ , \label{h00to2}
\end{eqnarray}
where
\begin{eqnarray}
 Q_0 =\frac{16\pi GC_0}{1-8\pi G\beta_{10}}\left(\beta_{10}+2\beta_{30} +\frac{1-8\pi G\beta_{10}}{8\pi
 G}B_0\right).
\end{eqnarray}
The last term of Eq.~(\ref{h00to2}) is a new PPN parameter.
However, there is no need to introduce any additional PPN
parameters when we move into the standard gauge by choosing $B_0$
such that $Q_0 =0$. We have
\begin{eqnarray}
 B_0 = -\frac{8\pi G(\beta_{10}+2\beta_{30})}{1-8\pi G\beta_{10}}.
\end{eqnarray}

We have all the necessary ingredients to read off the PPN
parameters. Let us begin by writing the full metric, we have
\begin{eqnarray}
g_{00} &=& -1 + 2U-2U^2 +4\Phi_2
\nonumber\\
&&+ 4\Phi_1 + 2\Phi_3 + 6\Phi_4  \ , \nonumber\\
g_{ij} &=& (1+2U)\delta_{ij} \ , \nonumber\\
g_{0i} &=&  \frac{\beta_{10}}{\beta_{10}-8\pi G\left(\beta_{10}^2-\beta_{30}^2 \right)} \times \nonumber\\
&& \left\{ \left[4(1-8\pi G\beta_{10})+16\pi GE_{0}-B_{0}C_{0} -
{1\over 2} \right]V_i \right. \nonumber\\
&&\left.- \left(16\pi GE_{0}-B_{0}C_{0} - {1\over 2}\right)W_i
\right\} \ . \label{eq:resultingmetric}
\end{eqnarray}
By comparing Eqs.~(\ref{eq:PPN metric}) and
(\ref{eq:resultingmetric}), the PPN parameters are given by
\begin{eqnarray}
&&\gamma_{PPN}=1,~ \beta_{PPN} = 1,~ \chi=\zeta_1=\zeta_2=\zeta_3=\zeta_4=\alpha_3=0, \nonumber\\
&&\left({1\over 8\pi G}\right)\alpha_1=-\frac{8\beta_{30}^2}{\beta_{10}(1-8\pi G\beta_{10})+8\pi G\beta_{30}^2} \ , \nonumber\\
&&\left({1\over 8\pi G}\right)\alpha_2=\frac{(\beta_{10}+2\beta_{30})^2}{(1-8\pi G\beta_{10})(\beta_{10}+\beta_{20}+\beta_{30})} \nonumber\\
&&-\frac{6\beta_{10}\beta_{30}+3\beta_{10}^2[1-8\pi
G(\beta_{10}+2\beta_{30})]}{(1-8\pi
G\beta_{10})\left[\beta_{10}(1-8\pi G\beta_{10})+8\pi
G\beta_{30}^2\right]}\nonumber\\
&&-\frac{4\beta_{30}^2[1-2\pi G(\beta_{10}-6\beta_{30})]}{(1-8\pi
G\beta_{10})\left[\beta_{10}(1-8\pi G\beta_{10})+8\pi
G\beta_{30}^2\right]}
 \ . \label{eq:resultppn}
\end{eqnarray}

Notice that the values of $\gamma_{PPN}$ and
$\beta_{PPN}$ are the same as in GR.  $\gamma_{PPN}=1$ implies an
identical predicted deflection of light about the Sun as well as
identical predictions for radar echo delay. The parameter
$\beta_{PPN}$ enters into the expression for anomalous
relativistic precession of planetary orbits, but the strongest
experimental limit is provided by the lunar laser ranging test.
The other non-vanishing ones are $\alpha_1$ and $\alpha_2$ which
are the effect of the preferred frame.

\section{Conclusions}
\label{sec:conclu} In this paper, we have investigated the
cosmological evolution of an interacting scalar field model in
which the scalar field has an interaction with the background
matter via Lorentz violation. We propose a model of interaction,
specifically $Q_m=- \dot{\bar{\beta}}\rho_m/\bar{\beta}$ in which
the interaction is mediated by the slope of the effective coupling
$\bar{\beta}$. The equation of state parameter of the scalar field
is expressed by eq.~(\ref{eq:eos-darkmatter}) as a candidate of
dark energy. The important role of the model is played by the
effective coupling in the transition era from the matter dominated
to scalar field dominated, which leads to an accelerating
universe. The model also predicts a constant fraction of dark
energy to dark matter in the future and hence solves the
coincidence problem. This is a profitable support to the effective
coupling. As a cosmological implication, the dynamic of the
effective gravitational constant is determined by the effective
coupling and allows one to test the Lorentz violating
scalar-vector-tensor theory of gravity using the SnIa data. We
have studied how a varying $G$ or a effective coupling could
modify the evolution of the Hubble parameter which is deviated by
the term of ${\bar{\beta}}^{-2}$. For a simple polynomial
$\bar{\beta}(z)=\bar{\beta}_0(1+\zeta z^2)$ ansatz, the best fit
values are $\chi^2_{min}=195.68$, $\zeta=-0.33$, and
$\Omega_{m0}=0.24$ for the modified $\Lambda$CDM model and
$\chi^2_{min}=195.71$, $\zeta=-0.29$, and $\omega_{\phi}=-1.13$
for the modified quintessence model.

We also have presented the 1PPN parameters of the theory. Our
result strongly depends on the two assumptions: the scalar field
satisfies the extremum condition, and the coupling functions
$\beta_1$ and  $\beta_2$ are constant. The first assumption stems
from the fact that the potential will play the role of an
effective cosmological constant if the theory is to account for
the late time accelerated expansion of the universe. The second
one is to obtain the real velocity of the gravitational waves. Up
to 1PPN approximation, important to note that it is impossible to
obtain $G_N=G_{cosmo}$, although $\beta_1$ and  $\beta_2$ are
constant. Here $G_{cosmo}=G^{(e)}$ is given by
Eq.~(\ref{gcosmology}). $G_{cosmo}$ is still dynamic in this case
because of $\beta_2$, while $G_N$ is always constant
$(\ref{defNewton})$. In the post-Newtonian approximation, the
effect of non-constant coupling function $\beta_2$ appears in
${\cal O}(2.5)$ or higher.

Of course, there are many remaining works to make this scenario
more concrete which is beyond the main aim of the present work.
For instance, non-linear coupling or more complicate functions are
also possible.  In the present work we have considered to the case
of instantaneous critical points, where $\lambda_1$ and
$\lambda_2$ are the constant parameters. However, in order to
construct viable model, $\lambda_1$ and $\lambda_2$ should satisfy
the condition Eq.~(\ref{def-gamma}). We need to solve the
dynamical system Eq.~(\ref{def-lambda}). Then, $\lambda_1$ and
$\lambda_2$ are dynamically changing quantities. For example, the
critical point $D$ in Table~\ref{table1} becomes
$x(N)=\sqrt{3/2}/(\lambda_1(N)+\lambda_2(N))$, $y(N)=x(N)$,
$z(N)=0$, and $u(N)=0$ where $N=\ln a$. As a consequence we obtain
the running critical point according to the changing of
$\lambda_1(N)$ and $\lambda_2(N)$.

Another important aspect is that the coupling interaction between
the scalar field and matter fields affects only the solution for
the scalar field. $\phi^{(1)}$ represents the local deviation from
$\phi_0$, which vanishes far from the local system. In conclusion,
one can reasonably state that the our gravity model could be a
viable candidate theory, even in the PPN approximation. It cannot
be a priori excluded at Solar system scales.

\section{Acknowledgements}
\label{sec:Acknowledg} Arianto wishes to acknowledge all members
of the Theoretical Physics Laboratory, the THEPI Division of the
Faculty of Mathematics and Natural Sciences, ITB, for the warmest
hospitality. We would like to thanks K. Yamamoto, Theoretical
Astrophysics Group, Hiroshima University, for useful discussion.
This work was supported by Hibah Kompetensi No.
223/SP2H/PP/DP2M/V/2009.

%

\end{document}